\documentclass[sigconf]{acmart}

\AtBeginDocument{%
  \providecommand\BibTeX{{%
    \normalfont B\kern-0.5em{\scshape i\kern-0.25em b}\kern-0.8em\TeX}}}

\setcopyright{rightsretained}
\copyrightyear{2018}

\usepackage{enumitem}

\acmConference{\vspace{2pt}Computer Science \& Engineering}{York University, Toronto}{2010}

\begin{document}

\title{Metrics for Multi-Touch Input Technologies}

\author{Ahmed Sabbir Arif}
\affiliation{%
  \institution{Department of Computer Science and Engineering}
  \institution{York University}
  \streetaddress{4700 Keele Street}
  \city{Toronto}
  \state{Ontario}
  \country{Canada}}
  \postcode{M3J1P3}
\email{asarif@cse.yorku.ca}

\begin{abstract}
Multi-touch input technologies are becoming popular with the increased interest in touchscreen- and touchpad-based devices. A great deal of work has been done on different multi-touch technologies, and researchers and practitioners are frequently coming up with new ones. However, it is almost impossible to compare such technologies due to the absence of multi-touch performance metrics. Designers usually use their own methods to report their techniques' performances. Moreover, multi-touch interaction was never modeled. That makes it impossible for designers to predict the performance of a new technology before developing it, costing them valuable time, effort, and money. This article discusses the necessity of having dedicated performance metrics and prediction model for multi-touch technologies, and ways of approaching that.
\end{abstract}

\keywords{Performance metrics, multi-touch input technologies, error correction, prediction, mathematical model.}

\begin{teaserfigure}
  \includegraphics[width=\textwidth]{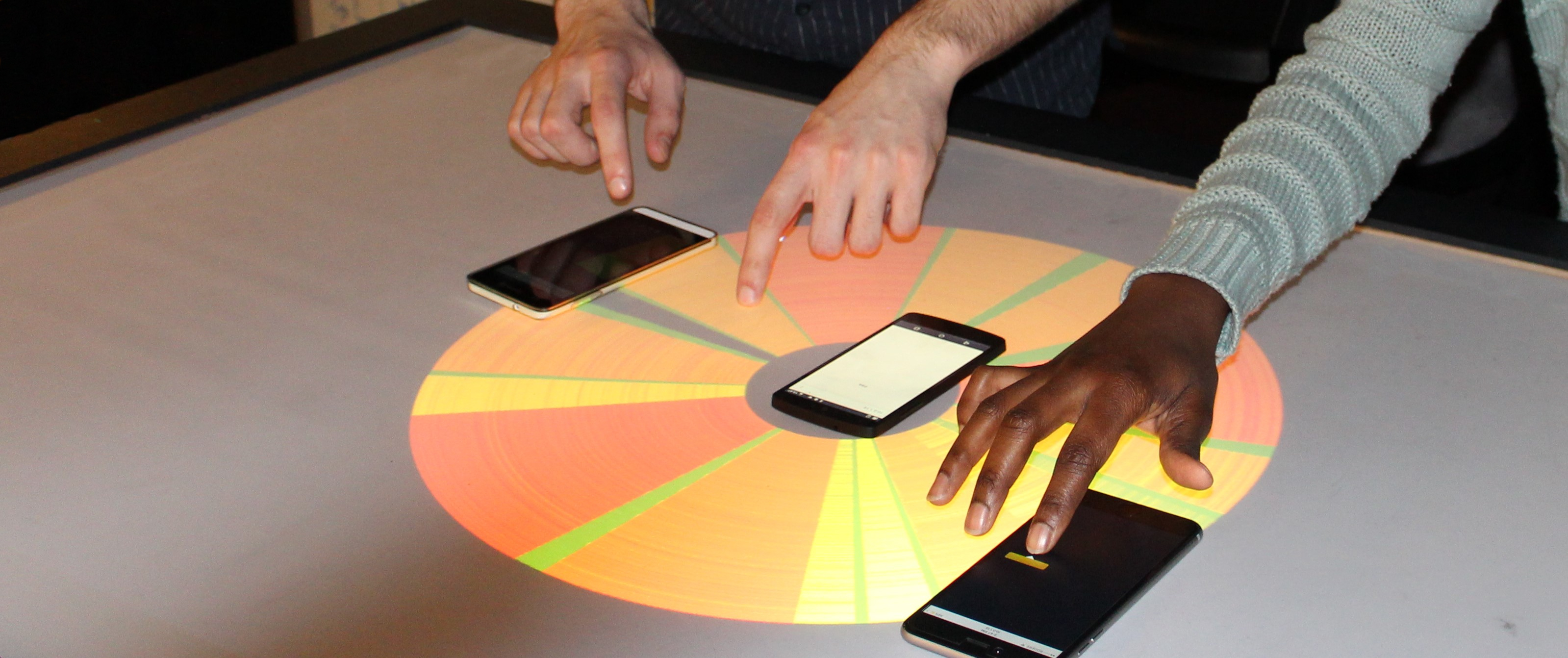}
  \caption{Multi-touch technologies are emerging as a major medium of high-degree of freedom interaction. Picture courtesy of Synlab.}
  \vspace{10pt}
  \label{fig:teaser}
\end{teaserfigure}

\maketitle

\section{Introduction}
Touchscreen- and touchpad-based multi-touch technologies are emerging as a major medium of high-degree of freedom interaction (Figure \ref{fig:teaser}). A great deal of work has been done on different multi-touch technologies, recognition algorithms, applications (e.g., \cite{benko_sphere_2008,davidson_synthesis_2006,dietz_diamondtouch_2001,echtler_shadow_2008,han_low-cost_2005,hodges_thinsight_2007,lee_multi-touch_1985,letessier_visual_2004,malik_interacting_2005,matejka_design_2009,chia_shen_multi-user_2006,von_hardenberg_bare-hand_2001,westerman_multi-touch_2001,zhang_visual_2001}), and researchers and practitioners are constantly coming up with new ones. But, it is almost impossible to compare these technologies due to the absence of multi-touch performance metrics. Designers usually use their own methods, which are typically modified versions of existing metrics, to report their techniques' performances. These metrics, however, fail to provide a clear picture of how the technologies work because of the \textit{direct} interaction strategies. Multi-touch techniques input directly to the device, under the points of contact such as, fingers, making it notably different from most other interaction technologies.

As most multi-touch metrics were coined by the designers to show how well their technologies perform, rather than to offer a good set of metrics, they are usually very straightforward and domain-specific. In other words, metrics used on a specific multi-touch technology cannot be used with any other technology. In addition to that, many assumptions were made to keep the metrics simple. For instance, some assume that all errors are exclusively due to the user, while other metrics assume it is the system that makes errors. In general, there is no differentiation between the human and system errors. Moreover, important human factors, such as finger and hand movements, cognitive processing and decision making times, and so on, as well as system factors, such as input processing time, flexibility, etc. are frequently overlooked.

To date, no attempt has been made on modeling the task of multi-touch interaction, either. Hence, it is impossible to predict a system's performance before developing it, causing designers to waste time, effort, and money. Multi-touch technologies have the potential of becoming one of the primary interaction techniques in near future, as touchscreen- and touchpad-based devices, such as tabletops, smartphones, tablets, etc. are rapidly emerging. At this stage it is imperative to have dedicated performance metrics and prediction model for such technologies. It will help us not only to compare novel techniques but also to predict the performance of new ones before implementing them.

This article starts with a brief discussion on current multitouch measurement techniques. It then, discusses the human and system factors that are likely indispensable when developing high-level multi-touch performance metrics and prediction model. Finally, it presents an outline of a potential future research.

\section{Current Metrics}
Almost all recent multi-touch empirical experiments report error rates along with other performance measures. In most cases, both performance and errors were classified by a straightforward \textit{hit-or-miss} strategy.

Participants were asked to perform specific tasks on a screen or a pad, and the experiment software kept a record of their actions. If the tasks were carried out successfully in a single attempt then it was considered a \textit{hit} or \textit{success}, if not then a \textit{miss} or an \textit{error}. Since different systems have different ways of interpreting user interaction hits and misses were counted differently based on the system design. One reason for this is that no research has been done on multi-touch metrics and error classification techniques. Hence, designers report performance in different ways. None of the methods differentiate between the human and system factors and overlook important factors, as pointed out previously.

\section{Developing New Metrics}
It is true that different interaction techniques have different ways of handling similar tasks. Yet, it is possible to develop domain independent performance metrics and prediction models by identifying high-level tasks that are common to all technologies. For instance, tasks such as selecting, moving, or rotating an object, are common to almost all multi-touch techniques. These high-level tasks, then, can be broken down into low-level domain-specific tasks. This strategy has been proven effective while modeling other interaction technologies \cite{arif_predicting_2010}.

The question, what high- and low-level parameters need to be considered in new metrics and prediction model requires careful study of current multi-touch technologies and a better understanding of the real-life user interactions. However, at this point we can include at least the following human and system factors.

\subsection{Human Factors}
The two parameters below are cognitive processing times that can be recorded via empirical studies. Alternatively, these values can be collected from existing work, as it is safe to assume that these cognitive pauses are fairly uniform in lengths \cite{kieras_using_1993}.
\begin{itemize}
  \item[$\square$] $T_{preparation}^h$ or the preparation time is the average time it takes to make the decision to perform a task.
  \vspace{4pt}
  \item[$\square$] $T_{verify}^h$ or the verification time is the average time it takes to verify correct completion of a performed task.
\end{itemize}
The parameter below is the physical movement time that can be calculated using Hick--Hyman and/or Fitts' law \cite{seow_information_2005}. The first law can be used to measure the choice reaction time and the latter to measure the rapid aimed movements.
\begin{itemize}
  \item[$\square$] $T_{move}^h$ or the movement time is the average time it takes to move fingers or hands from one location to another.
\end{itemize}
The parameter below is the probability of making an error while performing a task, which can be determined based on the average error rate measured in empirical studies.
\begin{itemize}
  \item[$\square$] $R_{error}^h$ or the human error rate is the average probability of making an error while performing a task.
\end{itemize}

\subsection{System Factors}
Although it is not possible to be definite about the behavior of first two parameters below without conducting empirical experiments, it can be assumed that the values of these will be the sum of a growing and a decaying series, respectively.
\begin{itemize}
  \item[$\square$] $R_{learn}^s$ or the learning rate is the average asymmetric learning effect for a specific technology that represents how fast users learn, or get used to, a system's interface, functionalities, or even bugs. This parameter can prove useful when comparing performance between expert and novice users. However, the value for $R_{learn}^s$ can be considered zero when participants are well-trained or had lots of prior experience with the system.
  \vspace{4pt}
  \item[$\square$] $R_{use}^s$ or the usability rate expresses how user performance decreases over time due to the system's complexity or ergonomic discomforts. This factor may be necessary as most direct input technologies are known to cause physical discomfort, such as fatigue, stress, occlusions from the user's hand, and so forth, during long term usage or instabilities \cite{ha_direct_2006,wigdor_living_2007}.
  \vspace{4pt}
  \item[$\square$] $T_{process}^s$ or the input processing time is the average time it takes to process a low-level task, such as a drag, pinch, display output, etc., by a specific technology.
  \vspace{4pt}
  \item[$\square$] $R_{error}^s$ or the system error rate is the average probability of a system error, such as a misrecognition or an interpretation error, for a specific technology.
\end{itemize}

\subsection{Compound Factors}
Prior sections provided a partial list of potential human and system factors. Here, we present two potential compound factors.
\begin{itemize}
  \item[$\square$] $R_{error}$ or the error rate is the average of the compound of the human and system error rates, in other words the relationship between $R_{error}^h$ and $R_{error}^s$.
  \vspace{4pt}
  \item[$\square$] $T_{task}$ or task completion time is the average of the compound of the human and system times, in other words the relationship between  $T_{preparation}^h$, $T_{move}^h$, $T_{process}^s$, and $T_{verify}^h$, to perform a task in a single attempt.
\end{itemize}

These two compound parameters can be used as new multi-touch performance metrics: $R_{error}$ for measuring error rates and $T_{task}$ for measuring the overall performance of a specific technology. To determine how to calculate these parameters also require further research.

However, more research and studies are necessary to compile a complete list of parameters. For example, it may be necessary to find answers to questions such as the effect of the presence or absence of tactile feedback, which tasks are hard due to human limitations, the effect of constraints of human hands, and how the size and proximity of the display affects performance. It is also essential to find more precise relationships between the human and system factors to create high-level metrics and a predictive model.

\section{An Example}
A high-level goal can be ``move object'' that is actually the combination of small operations: \textit{select object}, \textit{drag object}, and \textit{release object}. These operations, too, are combinations of smaller operations: \textit{prepare to perform a task} that is $T_{preparation}^h$, \textit{perform the task} that is the relationship
between $R_{error}$ and $T_{task}$, and verify the task that is $T_{verify}^h$. This is how all major operations can be broken down into smaller and basic operations. This also makes it possible to present a predictive model.

\section{Future Research}
This section presents a plan for developing new metrics to measure and predict the performance of multi-touch input and interaction methods.

\paragraph{Stage-1: Studying Multi-Touch Technologies} At this stage, one must study existing technologies from the literature, as well as examine some academic and commercial devices in real-life scenarios, for a better understanding of the technologies. The main purpose will be to identify common goals, tasks, trends, patterns, discomforts, mistakes, and confusions. This will help identifying various low- and high-level human and system factors, and the relationships between them.

\paragraph{Stage-2: Preliminary Metrics and Model} Here the target is to define a set of preliminary performance metrics and create a predictive model based on the findings of the first stage.

\paragraph{Stage-3: Pilot Studies} In this stage, a series of pilot studies will be conducted to determine if the proposed metrics give the right kind of results. If not, the metrics must be fine-tuned based on the study results. This will eventually lead the research to a final set of metrics and a model.

\paragraph{Stage-4: Empirical Studies} After deriving the final metrics and prediction model, a full-length empirical study is needed for further verification. At this stage it is also a good idea to examine if the new metrics and model can be extended to related input technologies such as bimanual interaction \cite{leganchuk_manual_1998}.

\section{Conclusion}
Well-defined performance metrics and prediction models are important for the continued development of any maturing technology. Multi-touch is maturing rapidly, with promising trends. Researchers and practitioners are coming up with new multi-touch techniques in regular basis. But it is almost impossible to compare, evaluate, or predict the performance of systems, as, to date, there is no standard metrics or model for multi-touch. This research plan presents one potential avenue to identify such metrics and models and also an outline of work that can build on it.

\balance{}

\bibliographystyle{ACM-Reference-Format}
\bibliography{refs}

\end{document}